\documentclass[
    ,final            % use final for the camera ready runs
  ]
  {aipproc}

\usepackage{amsmath}
\usepackage{hyperref}

\layoutstyle{8x11double}

\newcommand{\mhalf}[1]{\frac{#1}{2}}
\newcommand{\ka}{\kappa}

\newcommand{\de}{\delta} % delta
\newcommand{\ba}{\begin{align}}
\newcommand{\ea}{\end{align}}	%doesn't work, don't know why

\newcommand{\nnnl}{\nonumber\\}	%nonumber new line

\newcommand{\dhalf}{\frac{d}{2}} % d/2
\begin{document}

\title{Non-perturbative analysis of the Gribov-Zwanziger action}


\author{Markus Q. Huber\footnote{Speaker at the Quark Confinement and the Hadron Spectrum IX Conference, August 30 - September 3, 2010, Madrid}\,\,}{
  address={Physikalisch-Theoretisches Institut, Friedrich-Schiller-Universit\"at Jena, Max-Wien-Platz 1, 07743 Jena, Germany}
}

\author{Reinhard Alkofer}{
  address={Institut f\"ur Physik, Universit\"at Graz, Universit\"atsplatz 5, 8010 Graz, Austria}
}

\author{Silvio P. Sorella}{
  address={Departamento de F\'isica Te\'orica, Instituto de F\'isica, UERJ - Universidade do Estado do Rio de Janeiro, Rua S\~ao Francisco Xavier 524, 20550-013 Maracan\~a, Rio de Janeiro, Brazil}
}

\begin{abstract}
 In the non-perturbative regime the usual gauge fixing is not sufficient due to the Gribov problem. To deal with it one can restrict the integration in the path integral to the first Gribov region by using the Gribov-Zwanziger action. In its local form it features additional auxiliary fields which mix with the gluon at the two-point level. We present an explicit infrared analysis of this action. We show that from the two possible scaling solutions obtained previously only one remains: It coincides exactly with the results from the Faddeev-Popov action, i.e., the ghost propagator is infrared enhanced and the gluon propagator infrared suppressed and the corresponding power law behavior is described by only one parameter $\kappa=0.5953\ldots$. This corroborates the argument by Zwanziger that for functional equations it suffices to take into account the appropriate boundary conditions and no explicit restriction in the path integral measure is required.
\end{abstract}

\maketitle

Understanding confinement, i.e., the absence of quarks and gluons from the physical spectrum, has been a long-standing challenge for the theory of strong interations. In the course of time several seemingly different confinement scenarios have emerged. By now it has become clear that different explanations are not mutually exclusive but are only different manifestations. Especially in the last fifteen years functional methods have provided new insights into some of these scenarios, as the basic quantities in these approaches, the Green functions, give direct access to some of them, see, e.g., \cite{Alkofer:2000wg,Braun:2007bx}. Because of its accessibility mostly the Landau gauge has been investigated in this context, see, for example, \cite{vonSmekal:1997vx,Zwanziger:2001kw,Lerche:2002ep,Pawlowski:2003hq,Fischer:2008uz}.

A particular gluon confinement picture is due to Gribov and Zwanziger \cite{Gribov:1977wm,Zwanziger:1989mf}, where the gluon is removed from the physical spectrum due to a strong suppression of its propagator at low momenta. The argument is based on the insufficient gauge fixing implemented by the Faddeev-Popov procedure \cite{Gribov:1977wm}. While this is not harmful for perturbation theory it is important for the non-perturbative regime. The so-called Gribov-Zwanziger action incorporates a refined restriction in field configuration space in an attempt to avoid this problem \cite{Gribov:1977wm,Zwanziger:1989mf}: Instead of integrating over all configurations fulfilling the Landau gauge condition $\partial A=0$ one takes into account only configurations from the first Gribov region $\Omega$ defined as $\{A|\partial A=0, \, -\partial D>0\}$.
As a result already the tree level gluon propagator is confined because it vanishes at zero momentum: In this case positivity, a requirement for any physical particle, is violated maximally \cite{Zwanziger:1991gz}. The ghost propagator on the other hand is found to diverge more strongly than the tree-level part, namely like $1/p^4$ \cite{Gribov:1977wm,Zwanziger:1992qr}. This result is only obtained when enforcing the horizon condition, which arises as an additional condition in the gauge fixing procedure. Also in our approach it plays a pivotal role.

Although the insufficiency of the standard gauge fixing has been long known, Dyson-Schwinger equations (DSEs) relied up to now on the Faddeev-Popov action. To understand why this is working we refer to a conjecture of Zwanziger \cite{Zwanziger:2001kw,Zwanziger:2003cf}: Deriving DSEs with an integration in field configuration space formally restricted to the Gribov region does not change their form. The reason is that cutting the integration at the Gribov horizon does not introduce any boundary terms as the Faddeev-Popov determinant, which appears in the integrand of the path integral, vanishes on the horizon. Hence one can use the Faddeev-Popov action but has to take into account a proper boundary condition.

The original Gribov-Zwanziger action contains a non-local term \cite{Zwanziger:1989mf}, which can be localized with auxiliary fields $\bar{\omega}_\mu^{ab}$, $\bar{\omega}_\mu^{ab}$, $\bar{\varphi}_\mu^{ab}$ and $\bar{\varphi}_\mu^{ab}$ \cite{Zwanziger:1989mf}. The statistics obeyed by these fields is important: The first two are anti-commuting and the other two commuting. The additional adjoint color index accounts for the correct number of degrees of freedom. For our purposes the local action is rewritten as
\begin{align}\label{eq:final-GZ-action}
S&=\int dx \Bigg(\frac1{4}F^a_{\mu\nu}F^a_{\mu\nu}+\frac{1}{2\xi}(\partial_\mu A_\mu)^2-\bar{\eta}^{a}_c \,M^{ab}\, \eta^{b}_c+\nnnl
 &+\frac1{2} V_\mu^{ac}\,M^{ab}\,V_\mu^{bc} +i\, g\,\gamma^2 \sqrt{2} f^{abc} A_\mu^a V_\mu^{bc}\Bigg),
\end{align}
where the field $V_\mu^{ab}$ is the imaginary part of $\bar{\varphi}_\mu^{ab}$ and $\varphi_\mu^{ab}$:
\begin{align}
\varphi=\frac1{\sqrt{2}}\left( U+i\,V\right), \quad \bar{\varphi}=\frac1{\sqrt{2}}\left( U-i\,V\right).
\end{align}
The anti-commuting fields $\bar{\eta}_c^a$ and $\eta_c^a$ comprise the Faddeev-Popov ghosts, the real part of $\bar{\varphi}_\mu^{ab}$ and $\varphi_\mu^{ab}$ and the fields $\bar{\omega}_\mu^{ab}$ and $\bar{\omega}_\mu^{ab}$. The index $c$ runs from $\frac{d}{2}(N^2-1)+1$. Treating all these fields with only one pair of fields is possible, because they all interact only via the Faddeev-Popov operator $M=-\partial D$. The massive parameter $\gamma$ is the Gribov parameter. Its value is fixed by the horizon condition. Although the part proportional to $\gamma$ breaks BRST invariance, see, e.g., \cite{Zwanziger:1992qr,Capri:2010hb}, this action is renormalizable. See \cite{Dudal:2010fq} and references therein for more details on the renormalization and the horizon condition.

DSEs for this action can be derived with the \textit{Mathematica} package \textit{DoDSE} \cite{Alkofer:2008nt} which is of great help here, as due to the existence of mixed propagators even the two-point DSEs become quite complicated: The $AA$, $VV$, $\eta\bar{\eta}$, $VA$ and $AV$ two-point function DSEs have $37$, $5$, $3$, $5$ and  $36$ terms. The equations can be found in \cite{Huber:2009tx}.

For the analysis of the equations we write the full two-point function $\Gamma_{AA}:=\de^2\Gamma/\de A\de A$ as
\begin{align}
\Gamma^{AA,ac}_{\mu\nu}&=\delta^{ac}p^2 c_A^\bot(p^2) P_{\mu\nu}+\delta^{ac}\frac{1}{\xi} c_A^\parallel(p^2) p_\mu p_\nu.
\end{align}
For the two-point functions $\Gamma_{AV}:=\de^2\Gamma/\de A\de V$ and $\Gamma_{VV}:=\de^2\Gamma/\de V\de V$ we make ans\"atze which take only the tensors from the bare action into account:
\begin{align}
\Gamma^{VV,abcd}_{\mu\nu}&=\delta^{ac}\delta^{bd}p^2 c_V(p^2) g_{\mu\nu},\\
\Gamma^{AV,cab}_{\mu\nu}&=f^{cab} i\, p^2 c_{AV}(p^2) g_{\mu\nu}.
\end{align}
The functions $c_i$, $i=\{A, V, AV\}$, are non-perturbative dressing functions for the gluon, $V$ field and mixed $AV$ two-point functions.
In principle more tensors are possible. In order to obtain the corresponding propagators we have to invert the $(2\times2)$ two-point matrix $\Gamma^{\psi\psi}$, $\psi \in \{A, V\}$:
\begin{align}
&D^{AA,ab}_{\mu\nu}=\delta^{ab}\frac{1}{p^2} P_{\mu\nu} \frac{c_V(p^2)}{c_A^\bot(p^2) c_V(p^2)+2 N\, c^2_{AV}(p^2)},
\end{align}
\begin{align}
&D^{VV,abcd}_{\mu\nu}=\frac{1}{p^2}\frac{1}{c_V(p^2)}\delta^{ac}\delta^{bd}g_{\mu\nu}-\nnnl
 &\,-f^{abe}f^{cde}\frac1{p^2}P_{\mu\nu}\frac{2  c_{AV}^2(p^2)}{c_A^\bot(p^2) c_V^2(p^2)+2N\, c_{AV}^2(p^2) c_V(p^2)},\\
&D^{AV,abc}=-i\,f^{abc}\frac1{p^2}P_{\mu\nu}\frac{\sqrt{2} c_{AV}(p^2)}{c_A^\bot(p^2) c_V(p^2)+2N\,c_{AV}^2(p^2)}.
\end{align}
The propagator of the $\eta$ field is unaffected by the mixing:% and taken as
\begin{align}
D^{\eta\bar{\eta},ab}_{cd}=(\Gamma^{\eta\bar{\eta},ab}_{cd})^{-1}=-\de^{ab}\de_{cd}\frac{c_{\eta}(p^2)}{p^2}.
\end{align}

For determining the behavior of Green functions at low momenta we study the DSEs in this limit. Then the integrand is dominated by small momenta and all dressing functions can be replaced by their infrared (IR) expressions. For the dressing functions we make the ans\"atze $c_i(p^2)=d_i \cdot(p^2)^{\ka_i}$ and similarly for the vertices. The complete qualitative behavior is now described by the IR exponents $\ka_i$. For the propagators we cannot say which part will be dominant in the IR, so we have to distinguish four scenarios. For every case one can then perform an analysis along the lines of ref. \cite{Huber:2009wh}, where a general method for the calculation of possible IR scaling solutions is described. It was shown in ref. \cite{Huber:2009tx} that two scenarios are inconsistent, as the IR analyses yield conflicting conditions for the IR exponents. Hence we concluded that only two possible solutions remain, which are, however, both in qualitative agreement with the results from the Faddeev-Popov action and the IR leading diagrams are given by those which have a bare gluon-$\bar{\eta}\eta$ or gluon-$VV$ vertex. Here we will present evidence that one of those solutions can also be discarded.

This case was denoted III in \cite{Huber:2009tx} and is given by $\ka_A+\ka_V=2\ka_{AV}$, i.e., both parts of the determinant appearing in the propagators, $c_A^\bot(p^2) c_V(p^2)+2N\, c_{AV}^2(p^2) $, scale equally. The scaling analysis resulted in $\ka:=\ka_V=\ka_\eta>0$, $\ka_A=-2\ka+\dhalf-2$ and $\ka_{AV}=-\frac{\ka}{2}+\frac{d}{4}-1$, where $\ka$ is a constant to be determined. For this one has to solve a system of equations consisting of the four two-point DSEs truncated to their IR leading diagrams. The unknowns are the combinations of the dressing function constants $I_1=d_A d_\eta^2$, $I_2=d_{AV}^2 d_V$, $I_3=d_A d_V^2$ and $\ka$. As the determinant remains unchanged here, the variables appear in a more involved way than for the second case treated below, where the propagators have a simple dependence on the dressing functions. Consequently a complete analytic solution is not possible. However, one can solve for two variables analytically and try to find a solution numerically for the remaining equations. The required analytic expressions can be obtained with the current version of \textit{DoDSE} \cite{Alkofer:2008nt}. $I_2$ is then expressed analytically from one of the equations as is $I_1$ or $I_3$. The results are plugged into the remaining two equations. Finally, they are plotted for the allowed values of the variables: $\kappa$ is known to be positive from the scaling analysis and it has to be smaller than $1$ in order for the Fourier transformations of the propagators to exists. The values of $I_1$ and $I_3$ must be positive as the dressings $d_A$ and $d_\eta$ are necessarily positive. It turns out that no solution exists. Hence this case is ruled out for the chosen ansatz.

Therefore the only solution in our truncation is given by $\ka:=\ka_V=\ka_\eta>0$, $\ka_A=-2\ka+\dhalf-2$ and $\ka_{AV}>-\ka+\dhalf-2$ and the dominant part in the determinant is $c_A^\bot(p^2) c_V(p^2)$. However, for the $V$ propagator the second term is suppressed compared to the first one, so that its final form in the IR is
\begin{align}
D^{VV,abcd}_{\mu\nu}=\frac{1}{p^2}\frac{1}{c_V(p^2)}\delta^{ac}\delta^{bd}g_{\mu\nu}.
\end{align}

As a consequence the dependence of the IR leading diagrams in the two-point DSEs of the $V$ and $A$ field on the mixed propagator vanishes. Even more, the equations turn out to be completely the same in the IR as those obtained from the Faddeev-Popov action, because the contributions from the $V$ field and the $\eta$ fields cancel each other such that only contributions equivalent to those from the Faddeev-Popov ghosts remain. Hence also the result for $\ka$, which can be calculated analytically, is the same: $0.5953\ldots$ \cite{Zwanziger:2001kw,Lerche:2002ep}. Furthermore, the IR exponents of all vertices are determined as $\ka_{2n,m}=(n-m)\ka+(1-n)\left(\mhalf{d}-2\right)$ \cite{Alkofer:2004it,Huber:2007kc}. With this information the IR exponent of the mixed two-point function can be calculated. It turns out that there are several possible solutions for $\ka_{AV}$, the smallest one being $0.0668776$.

We want to stress that the solution obtained here solves the complete tower of DSEs in the IR and is the only possible solution. An important point is how the horizon condition enters. Without it we only could get the trivial result of canonical scaling of all Green function because of the bare two-point functions appearing on the right-hand sides of the two-point DSEs. However, solving the DSEs requires a renormalization condition which can be chosen as the value of the ghost dressing function at zero momentum. It was shown in ref. \cite{Fischer:2008uz} that this value decides between the scaling solution (when it is chosen to be 0) and a family of decoupling solutions. Here the horizon condition explicitly forces us to use the former renormalization \cite{Zwanziger:2001kw,Gribov:1977wm,Zwanziger:1992qr}. This is how this additional condition enters and why we get the scaling solution and no decoupling type of solution like, for example, in \cite{Fischer:2008uz,Aguilar:2008xm,Alkofer:2008jy}. In principle the analysis also allows for other possibilities, e.g., for an IR divergent gluon propagator, but they are not realized.
Discussions about the different solutions can be found, e.g., in \cite{Fischer:2008uz,Alkofer:2008jy,Sternbeck:2008mv,Maas:2009se}.

Our results are in qualitative agreement with those obtained in refs. \cite{Gracey:2009mj,Zwanziger:2010iz}, where some propagators of the auxiliary fields are found to diverge like $1/p^4$. This corresponds to the value $\ka=1$, which was already obtained by Gribov \cite{Gribov:1977wm} in a perturbative analysis. The agreement we find in the IR with the scaling solution of the Faddeev-Popov action is another indication of the solidity of this confinement scenario, since it corroborates the conjecture of Zwanziger \cite{Zwanziger:2001kw,Zwanziger:2003cf} about cutting the integration at the Gribov horizon.

Acknowledgments: Discussions with Daniel Zwanziger are gratefully acknowledged. M. Q. H. was supported by DFG projects Gi 328/1-4 and GRK 1523 and the FWF project W1203.

\bibliographystyle{aipproc}   % if natbib is available

\bibliography{literature_confinementIX}

\end{document}